\documentclass[11pt]{article}
\usepackage{graphicx}
\usepackage{natbib}
\usepackage{jaa}
\usepackage{times}

\begin{document}
\title{HD 12098 and Other Results from Naini~Tal~-~Cape Survey}
\author[V. Girish]
{V. Girish\thanks{e-mail: giri@tifr.res.in} \\
	Tata Institute of Fundamental Research, Homi Bhabha Road, Mumbai -
	400 005, INDIA \\
}
\maketitle

\begin{abstract}

Naini~Tal~-~Cape Survey is a survey started with the aim of searching for new
rapidly oscillating Ap stars in the northern hemisphere and has discovered
one mono-periodic roAp star HD 12098. The frequency separation of HD~12098
suggests a rotation period of 5.5 day for the star.
The discovery of roAp oscillations in HD~12098 and the results of the
multi-site observation campaign organized to resolve the ambiguity in the
determination of the rotation period of HD~12098 is presented.  The results
of  non oscillating Ap stars discovered in the survey and two promising roAp
candidates HD~17431 and HD~207561 are also presented. If confirmed, the
variability in HD~207561 will make it the first Am star showing roAP type 
rapid variability.

\end{abstract}

\section{Introduction}

Naini~Tal~-~Cape Survey is a  collaborative program between India and
South~Africa,  initiated with the main aim of searching for new
northern hemisphere rapidly oscillating Ap stars (roAp) and to study
them \citep{seetha01}.  RoAp stars are short period photometric variables
discovered in 1978 by \Citet{kurtz78}. The period of oscillations in these stars
lie in the range of 5-21~minutes with typical amplitudes of few
milli~magnitude. The low amplitude of oscillations and the short periods
demand a very stable site and atleast one meter class telescope for the
study of roAp stars. For these reasons, the 104 cm Sampurnanand telescope
at ARIES, Naini~Tal is selected for the survey.  For a summary of
Naini~Tal~-~Cape Survey, and the site characteristics and facilities
available at ARIES, Naini~Tal for variable star research, readers are
referred to, \Citet{santosh05} and \Citet{sagar05}.

Naini~Tal~-~Cape Survey is an ongoing program. Till 2004, we have searched for short period variability in a total of 63
stars and detected rapid oscillations in HD 12098 \Citep{girish01},
$\delta$~Scuti oscillation in four stars \Citep{santosh05}.  The selection
of the candidate stars for the survey is made mainly on the basis of their Ap
nature, Str\"omgren colours, and temperature when ever available.  On the
basis of known roAp stars, \Citet{peter95} found that, roAp stars occupy
specific Str\"omgren color space. To increase the chances of detecting new
roAp stars, the candidates for the survey are  selected mainly on these
empirical limits on their Str\"omgren colours. Since the limits are only
empirical, the selection criteria is relaxed so as to include those stars
which might lie just outside these limits. In fact one of the six
Str\"omgren colors ($\delta m1$) of HD~12098, the first roAp star
discovered under the survey  fall outside the corresponding limit.

The Ap stars which share similar properties as that of roAp stars, but do
not show rapid variability are called as non-oscillating Ap (noAp) stars.
The noAp stars hold equal importance as that of roAp stars for the
understanding of roAp phenomenon.  During Naini~Tal~-~Cape Survey, we found
that three stars HD 25499, HD 38143 and HD 38817  fall under the category
of noAp stars.  The amplitude of variability in these three stars, if any,
is less than 0.2 mmag in the frequency range of 1-5mHz. In addition, we
observed roAp variability in HD~17431 and HD~207561 on few nights, but the
data at hand are not sufficient  to confirm the variability in these stars.
These results are discussed in the following sections.

\section{HD~12098}

HD 12098 is the first roAp star discovered in  Naini~Tal~-~Cape Survey.
The star is a bright ($m_V = 7.9$) F0 star. The Str\"omgren color indices
of the star ($b-y = 0.191,\ m_1 = 0.328,\ c_1 = 0.517 \ \mathrm{and}\ \beta
= 2.796$, \Citealt{hauck98}) combined with the de-reddened parameters
estimated using the calibrations by \Citet{crawford75} falls within the
empirical limits suggested for roAp stars  except for $\delta m_1$. The
discovery of roAp oscillations in HD~12098, thus extends the limits on
$\delta m_1$ suggested by \Citet{peter95} for probable roAp stars by a
slight margin.

\begin{figure}[b]
	\centering
	\includegraphics[width=1.2in,height=4.2in,angle=-90]{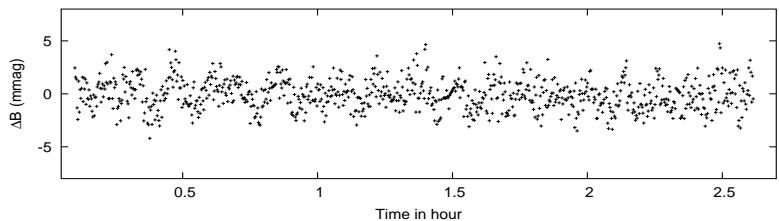}
	\caption{ Discovery lightcurve of HD~12098 observed on 1999 November 21.}
	\label{hd-discovery}
\end{figure}

On the basis of Str\"omgren colors, HD~12098 was selected as a candidate
for the survey  and observed on the night of 1999 November 21.
The discovery light curve plotted in Fig~\ref{hd-discovery} clearly shows
variability around 7.1~minutes.

The discovery and followup observations showed that the oscillation
amplitude in HD~12098 vary from night to night, indicating plausible
multi-mode oscillation and/or rotational modulation.  To determine the
nature of oscillations, HD~12098 was observed on six nights from
Gurushikhar observatory, Mt.~Abu and ARIES, NainiTal for a total of
65~hours.

\begin{figure}
	\begin{center}
	\includegraphics[height=3.5in,width=4in]{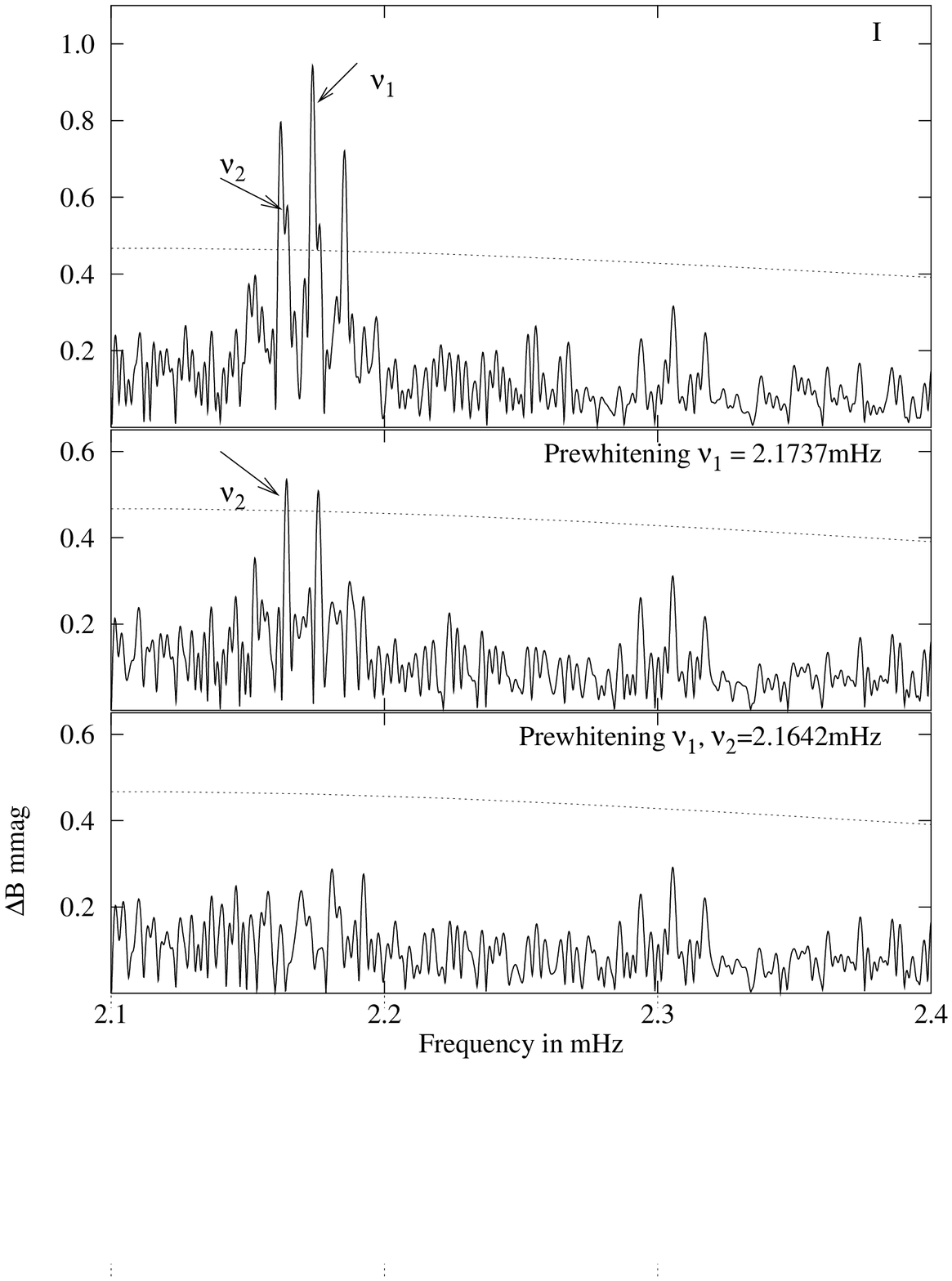}
\end{center}
	\caption{ Amplitude spectra of the data obtained from Mt.Abu merged
	together. See text for details}
	\label{fig-mtabu}
\end{figure}

The analysis of individual nights' data confirm the modulation in the
amplitude of oscillation. The amplitude modulation in roAp stars is caused
mainly by two processes. First by beating of close spaced periods and
second  due to the rotation of the star \Citep{kurtz82}.  To identify the
nature of oscillations in HD~12098,  the data obtained from Mt.Abu was
combined on a common time scale and subjected to discrete Fourier transform
(DFT) \citep{deeming75}.  The significance level of the periods in the
amplitude spectrum was determined on the basis of local noise.  The running
average of amplitudes of group of fifty frequencies are computed as the
local noise while treating all the peaks in the data are due to noise.  The
significance level was set at four times the local noise level
\Citep[see,][]{breger96}.  Any peak lying above this limit was treated
as real.  In Figure~\ref{fig-mtabu} this significance limit is shown in the
dotted  line.  The oscillation around 7.1~minute (2.173~mHz) well above the
significance limit is clearly seen in the Figure.


In Figure~\ref{fig-mtabu}, an unresolved frequency very close to the alias
peak (marked $\nu_2$) was clearly seen. To recover this frequency and to
check for any other peaks that might be buried under the main frequency at
$\nu_1$, a noise free sinusoid with the amplitude and frequency
corresponding to  $\nu_1$ was subtracted from the time series
(prewhitening). The residual data was subjected to DFT again. The resulting
amplitude spectra is plotted in the second panel of Fig.\ref{fig-mtabu}.
The spectrum show two peaks at $2.1641\ \mathrm{\&}\ 2.1759$~mHz with
similar amplitude. The frequency difference between the two frequencies
suggested one of them to be the alias of the second frequency.  Typically
the real frequency will have a higher amplitude and the aliases will be of
lower amplitude. But, for amplitudes close to the noise level, this may not
be true, making it  difficult to differentiate the real peak from the
alias.


\Citet{girish01} interpreted the second frequency as the rotationally split
component of the main frequency and predict a rotation period of either
1.22 or 5.5 day for HD~12098. The ambiguity was due to the ambiguity in the
identification of the second frequency. The magnetic field measurements of
HD~12098, however favour the longer rotation period \Citep{wade01}.


\subsection*{Multi-site campaign }

The single site observations of HD~12098 showed the second frequency as
the rotationally split component of the oscillation frequency of HD~12098.
However, the second frequency suffered  from 1 cycle/day alias ambiguity.

With the main aim of resolving the ambiguity in identifying  the second
frequency, and hence, rotation period of the star,  a multi-site
observation campaign on HD~12098 was organized in October/November 2002.
The campaign involved a total of eleven observatories,  with ten
observatories finally contributing to the data.  A total of 394 hours of
useful data extending over 28 nights with 45\% duty cycle were obtained.

From the analysis of the multi-site data,  mainly five frequencies
separated by $\delta\nu = 2.14 \pm 0.02$~mHz were obtained.  The
plausibility of such  small difference arising due to independently
excited modes was already ruled out \Citep{girish01}.
Assuming the frequency separation to be due to rotational splitting, the
equal separation of the frequencies gives a rotation period of $\Omega =
5.41 \pm 0.05$~days for HD~12098, very close to the 5.5~days rotation period
predicted by \Citet{girish01} assuming $2.1759$~mHz as the real
frequency instead of $2.1641 mHz$.  The amplitude modulation of
oscillations of individual nights' data also favours a period close to
$5.41$~days.  Detailed observation and analysis results of the multi-site
campaign will be published elsewhere \Citep{seetha05}.




\section{ Non oscillating Ap stars }

Non oscillating Ap stars hold equal importance in the understanding of roAp
phenomena. The systematic difference between the noAp stars and roAp stars
can help us in the understanding of possible reasons for the  incidence of
oscillations in roAp stars.  During Naini~Tal~-~Cape Survey survey, three
Ap stars were found to fit in to noAp star category.  The three noAp stars
HD~25499, HD~38143 and HD~38817 are discussed briefly.

\subsection{ HD~25499  }

\begin{figure}[htp]
	\centering
	\includegraphics[angle=-90,scale=0.40]{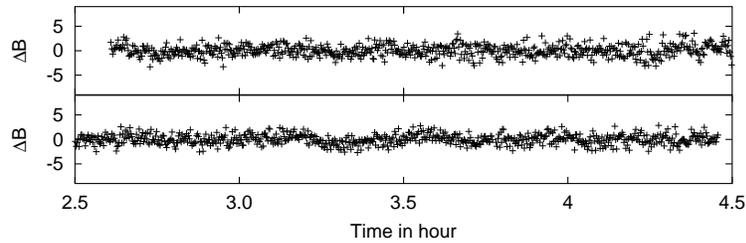}
	\caption{Lightcurve of HD~25499 observed on the night of 2003 November 10.}
	\label{fig-25499lc}
\end{figure}

\begin{figure}[htp]
	\begin{center}
	\includegraphics[angle=-90,scale=0.35]{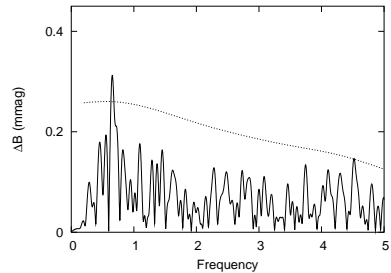}
\end{center}
	\caption{Fourier transform of the light curve plotted in
	Figure~\ref{fig-25499lc}. The dashed curve represents
	significance limit at three times local noise level.}
	\label{fig-25499ft}
\end{figure}

HD~25499 satisfies all the six Str\"omgren colors  used for identifying
probable roAp stars. The star was observed on six occasions for a total of
nine hours and found to show no variability at an amplitude limit of 0.2
mmag in the frequency range  $1 - 5$~mHz. The amplitude of oscillation, if
any would be below $0.2$~mmag at $3\sigma$ confidence level. The lightcurve
and the amplitude spectrum of HD~25499 observed on 10 Nov 2003 is plotted
in Figure~\ref{fig-25499lc} and Figure~\ref{fig-25499ft} respectively. Though,
there seems to be a peak above the significance level around 0.6~mHz
corresponding to 28~minutes, it is not  considered real due to the absence
of similar period in the other observations.


\subsection{HD~38143}

\begin{figure}[htp]
	\begin{center}
	\includegraphics[angle=-90,scale=0.35]{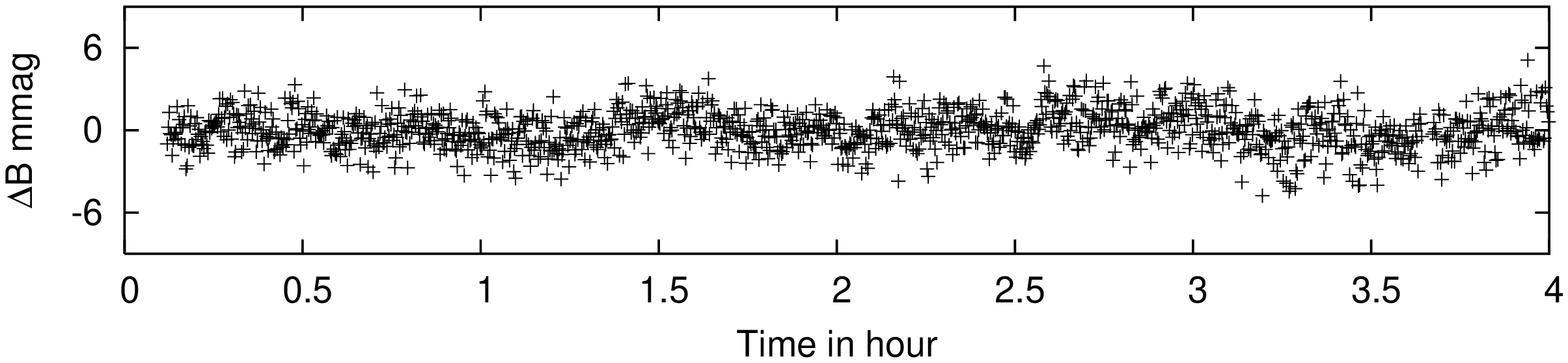}
\end{center}
	\caption{Lightcurve of HD~38143 observed on  2003November 19.}
	\label{fig-38143lc}
	\begin{center}
	\includegraphics[angle=-90,scale=0.40]{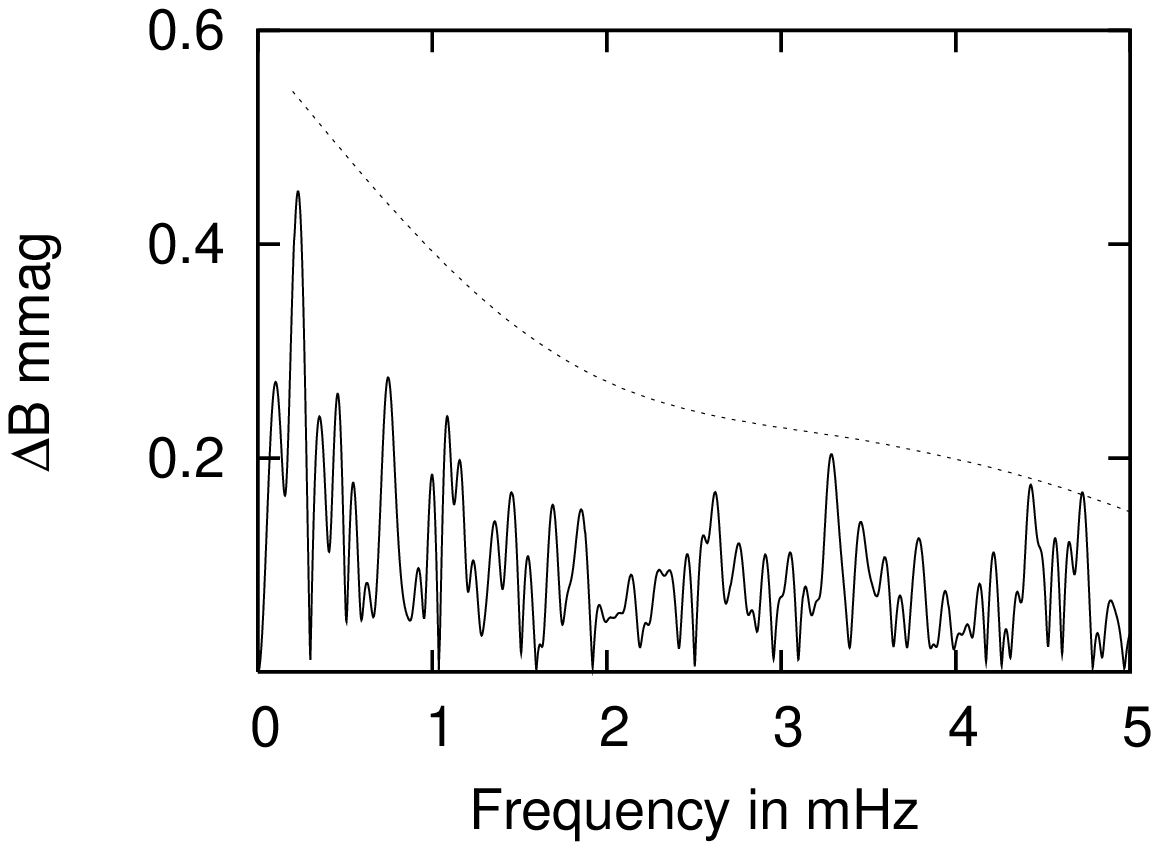}
	\end{center}
	\caption{Amplitude spectra of lightcurve plotted in
	Figure~\ref{fig-38143lc}}
	\label{fig-38143ft}
\end{figure}

HD~38143 is an A2 star whose Str\"omgren color indices lie well within the
empirical limits suggested by \Citet{peter95}. The star was first observed
on 23 January 2000, on three nights in 2002 and two nights in 2003 for a
total of 22 hours. No variability was seen in all the runs at an amplitude
limit of $0.3-0.4$~mmag. The lightcurve of HD~38143 observed on 2003
November 19 is  plotted in Figure~\ref{fig-38143lc} with the corresponding
amplitude spectrum in Figure~\ref{fig-38143ft}. The available observations
on HD~38143 suggest that the star is a noAp star, unless, HD~38143 has a
very long rotation period or the oscillations if any have amplitudes  below
our detection limit ($<0.3 mmag$).

\vspace*{-0.1in}
%

\subsection{ HD~38817 }

\begin{figure}[htbp]
	\begin{center}
	\includegraphics[angle=-90,scale=0.6]{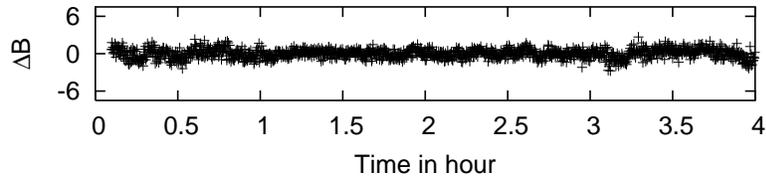}
\end{center}
	\caption{Lightcurve of HD~38817 observed on the night of 09 Nov 2003.}
	\label{fig-38817lc}
\end{figure}

\begin{figure}[htbp]
	\begin{center}
	\includegraphics[angle=-90,scale=0.5]{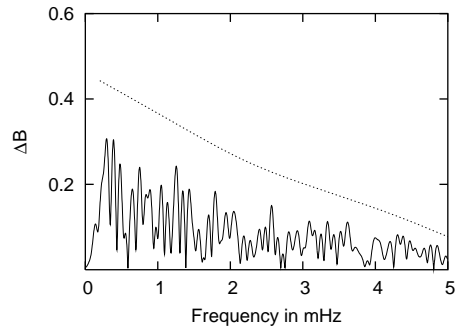}
\end{center}
	\caption{Plot of amplitude spectrum of the lightcurve plotted in
	Fig.\ref{fig-38817lc}}
	\label{fig-38817ft}

\end{figure}

HD~38817 is classified as an A2 star. Only three  of the six Str\"omgren
colors ($H_\beta$, $m_1$, $dm_1$) satisfy the limits suggested for a
probable roAp star, while $c_1$ lies with in 10\% of the
corresponding limits. The star was selected as a candidate star for
Naini~Tal~-~Cape Survey to check the possible extension of these limits,
but turned out to be a non-variable at a limiting amplitude of 0.2 mmag.
The star was observed for a total of 14 hours over four nights with the last
three nights' observations extending for more than three hours each. The
lightcurve and the corresponding amplitude spectrum of HD~38817 observed on
2003 November 09 are plotted in Figure~\ref{fig-38817lc} \&
\ref{fig-38817ft} respectively. From the analysis of the data we conclude
that HD~38817 is a noAp star at a limiting amplitude of 0.4~mmag at
$3\sigma$ confidence level.

\section{ Probable roAp star candidates }

In the course of our survey, we observed variability in the frequency range
of 1-5 mHz in two stars HD~207561 and HD~17431. But, the data on hand are
not sufficient to confirm variability in these stars. HD~207561 is
classified as an Am star \citep{floquet75} and the confirmation of rapid
variability will make it the first Am star showing roAp type variability.
For the benefit of the others who may be interested in these stars, we
briefly describe them.

\subsection{ HD~207561 }

\begin{figure}[ht] \begin{center}
	\includegraphics[height=5.2in,angle=-90]{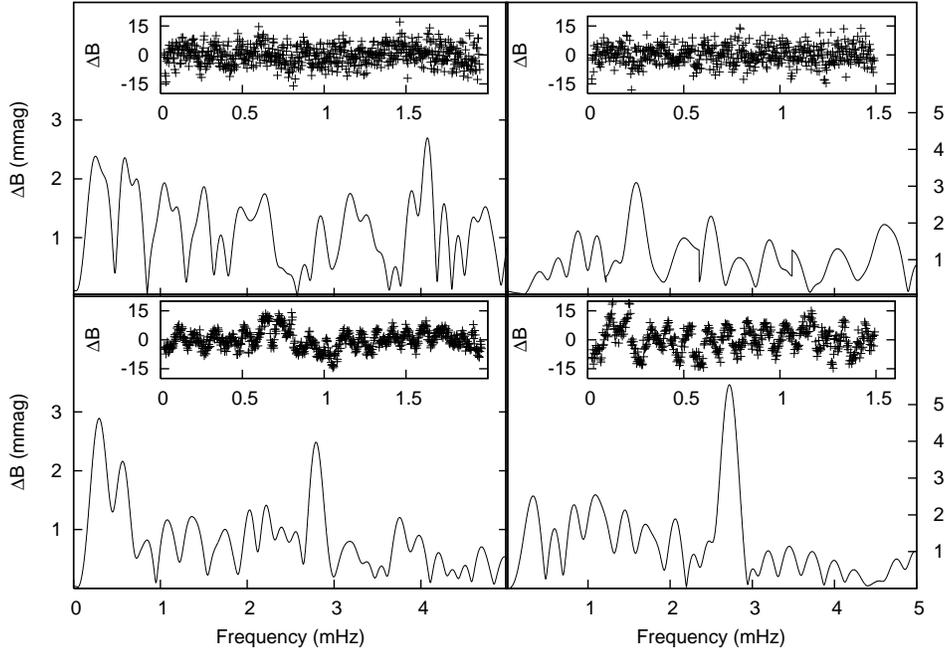}
	\caption[Amplitude spectrum of HD~207561]{Lightcurves and amplitude
	spectrum of HD~207561 observed on the nights of 6th Dec 2000 (left
	panels) and 07 Dec 2000 (right panels).  The top two panels shows the
	plots of the comparison stars while the bottom panels corresponds to
	HD~207561.  A peak around 6.1 minute period is clearly seen in
	both amplitude spectrum and lightcurve of HD~207561 which is absent in
	the comparison star data.}
	\label{fig-207561}
\end{center}
\end{figure}

HD~207561 is an F0~III star whose Str\"omgren colours fall well within the
empirical limits for probable roAp stars. The star was observed on 2000
December 16 for the first time.  The lightcurve and the corresponding
Fourier spectrum of HD~207561 obtained on two consecutive nights 2002
December 6 \& 7 are plotted in Figure~\ref{fig-207561} clearly show regular
variations.  Similar variations were absent in the comparison star data
observed simultaneously with the second channel.

Both the lightcurve and the amplitude spectrum  of HD~207561 plotted in
Fig.\ref{fig-207561} clearly show regular variability around six minutes
on two different nights.  The absence of similar variation in the
comparison star rules out the possibility of local effects. But the absence
of six minute period in the followup observations do not allow us to confirm
rapid oscillations in the star. Also the period is very close to six
minutes. When ever periodicity of integral multiple of a minute is
detected, care should be taken to rule out the plausibility of drive error
or other instrumental effects mimicking regular variability. Another likely
reason for the non-detection of oscillations in the follow up nights might
be due to mis-identification of the star. However, the confirmation of the
presence or absence of oscillations in HD~207561 will be important as it
will make HD~207561 the first Am star with roAp type variability.

\subsection{ HD~17431 }

\begin{figure}[htb]
	\hspace*{0.2in}
	\includegraphics[scale=0.21,angle=-90]{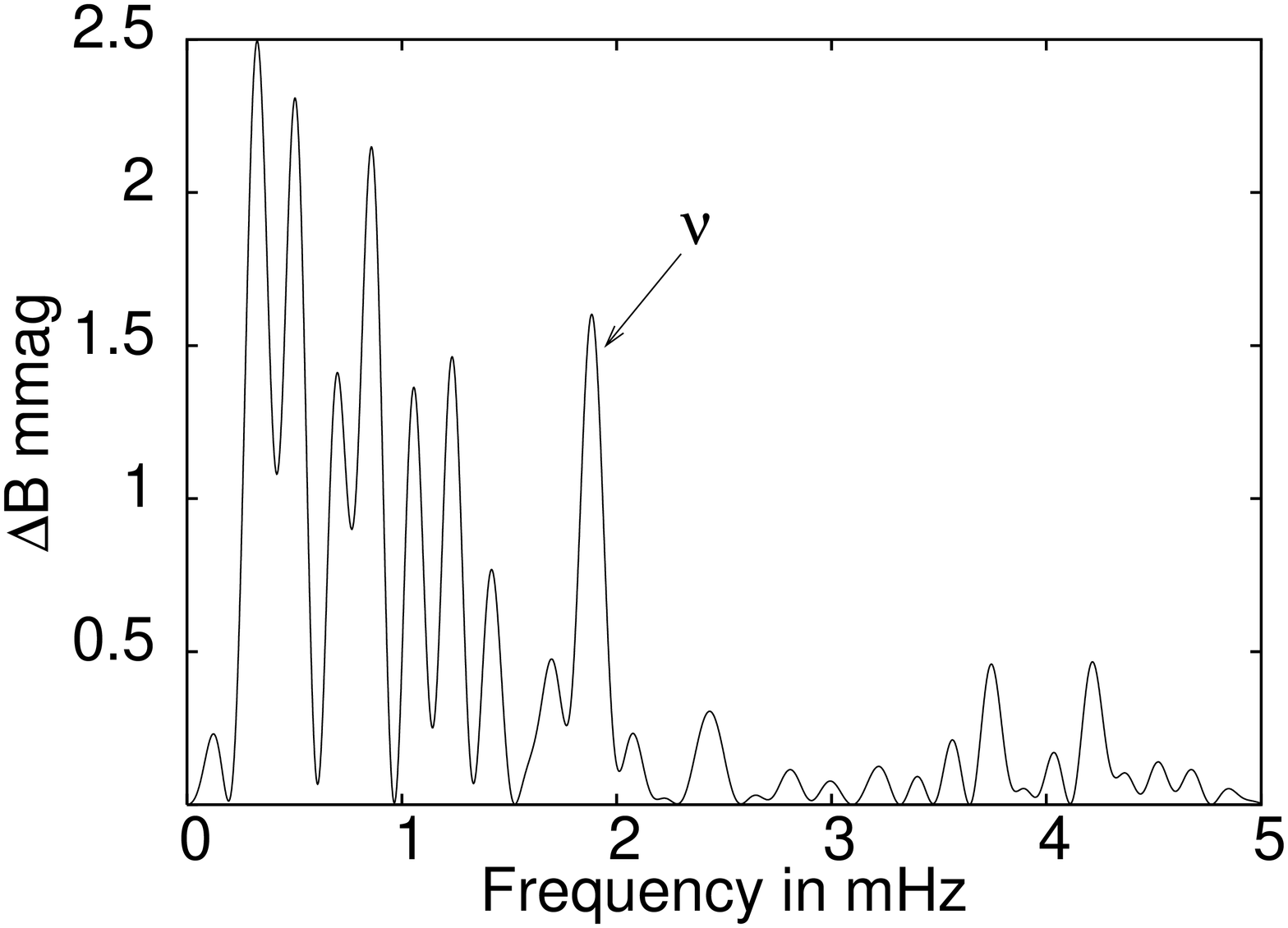}
	\includegraphics[scale=0.21,angle=-90]{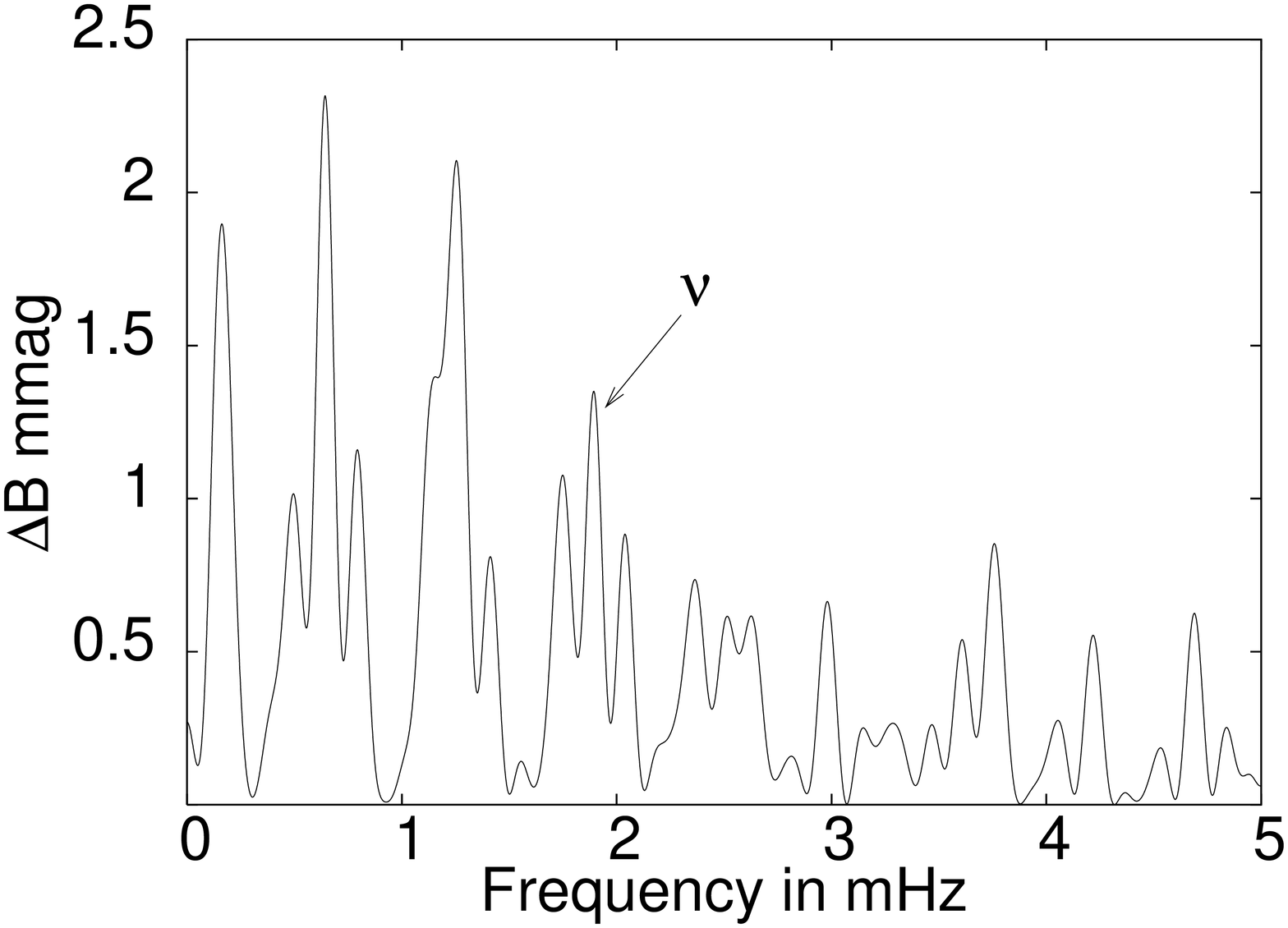}
	\caption{Amplitude spectrum of HD~17431 observed on 10th October 2000
	and 6th December 2001. The peaks marked as $\nu$ corresponds to $\sim
	8.8$ minutes.}
	\label{fig-17431}
\end{figure}

HD~17431 is classified as an A3 star. The Str\"omgren color indices of the
star satisfy all the six Str\"omgren parameters suggested for a probable
roAp star. The star was first observed on the night of 2000 October 2000
09. The Fourier transform of the data showed a peak around 8.8 minutes.
Similar periods were observed in the follow up observations though not with
very good signal except, on 2000 October 10. In Figure~\ref{fig-17431}, the
amplitude spectrum of the data obtained on 2000 October 10 and
2001 December 6 are plotted, which indicate the presence of a period around 8.8
minutes.  Though, similar variability near this period was seen on other
nights, the nights were not photometric and hence, prevents from confirming
variability in HD~17431.

\section{ Conclusions }

We present a brief summary of the main results of Naini~Tal~-~Cape Survey
conducted from late 1999 to early 2004. During this period one roAp star
HD~12098 a mono periodic oscillator was discovered. The amplitude
modulation of the pulsation frequency and frequency splitting observed in
HD~12098 suggest a rotation period of $\sim 5.4$~days for the star.  The
survey also resulted in the classification of three stars HD~25499,
HD~38143 and HD~38817 as non oscillating Ap stars,  useful in the
understanding of the differences between roAp and Ap stars sharing similar
properties of roAp stars but no oscillations. Also presented were two
promising roAp candidates HD~17431 and HD~207561 which showed variability
in few observations.  Further observations on the two are needed to confirm
variability in these stars. The confirmation of oscillations in HD~207561
will be an exciting result as it will make HD~207561 the first Am star  to
show rapid variability.

\section*{ Acknowledgements }

The author would like to thank Dr. S. Seetha for introducing
the exciting field of roAp stars.  She is also instrumental in
organizing  the multi-site campaign on HD~12098.  Thanks are due to Prof.
Ram Sagar for all the support at ARIES, NainiTal. The author would like to
acknowledge the help and support of Dr. Santosh and all the observatory
staff of ARIES, Naini~Tal with out whose help this survey would not be
possible.

\end{document}